\documentclass[lettersize,journal]{IEEEtran}
\usepackage{amsmath,amsfonts}
\usepackage{algorithmic}
\usepackage{array}
\usepackage[caption=false,font=normalsize,labelfont=sf,textfont=sf]{subfig}
\usepackage{textcomp}
\usepackage{stfloats}
\usepackage{url}
\usepackage{verbatim}
\usepackage{graphicx}
\usepackage{cite}
\usepackage{float}
\def\BibTeX{{\rm B\kern-.05em{\sc i\kern-.025em b}\kern-.08em
    T\kern-.1667em\lower.7ex\hbox{E}\kern-.125emX}}
\usepackage{balance}
\begin{document}

\raggedbottom

\title{A Quantitative Analysis of Physical Security and Path Loss with Frequency for IBOB Channel}

\author{Arunashish Datta, ~\IEEEmembership{Student Member,~IEEE,} Mayukh Nath, ~\IEEEmembership{Student Member,~IEEE,} Baibhab Chatterjee,~\IEEEmembership{Student Member,~IEEE,} Shovan Maity, ~\IEEEmembership{Member,~IEEE,} Shreyas Sen, ~\IEEEmembership{Senior Member,~IEEE}\vspace{-2em}
\thanks{This work was supported by the National Science Foundation (NSF) Career Award under Grant CCSS 1944602. The authors wish to acknowledge the support of the USDA, through the NRI and CPS programs, under grants  2018-67007-28439 and 2019-67021-28990.\\ The authors are with the School of Electrical and Computer Engineering, Purdue University, West Lafayette, IN 47907 USA. Corresponding author: Professor Shreyas Sen, e-mail: shreyas@purdue.edu }
}

\maketitle
%
%

\begin{abstract}
Security vulnerabilities demonstrated in implantable medical devices have opened the door for research into physically secure and low power communication methodologies. In this study, we perform a comparative analysis of commonly used ISM frequency bands and Human Body Communication for data transfer from in-body to out-of-body. We develop a Figure of Merit (FoM) which comprises of the critical parameters to quantitatively compare the communication methodologies. We perform FEM based simulations and experiments to validate the FoM developed.

\end{abstract}
%
%
\section{Introduction}

\begin{figure}[t!]
\centering
\includegraphics[width=0.45\textwidth]{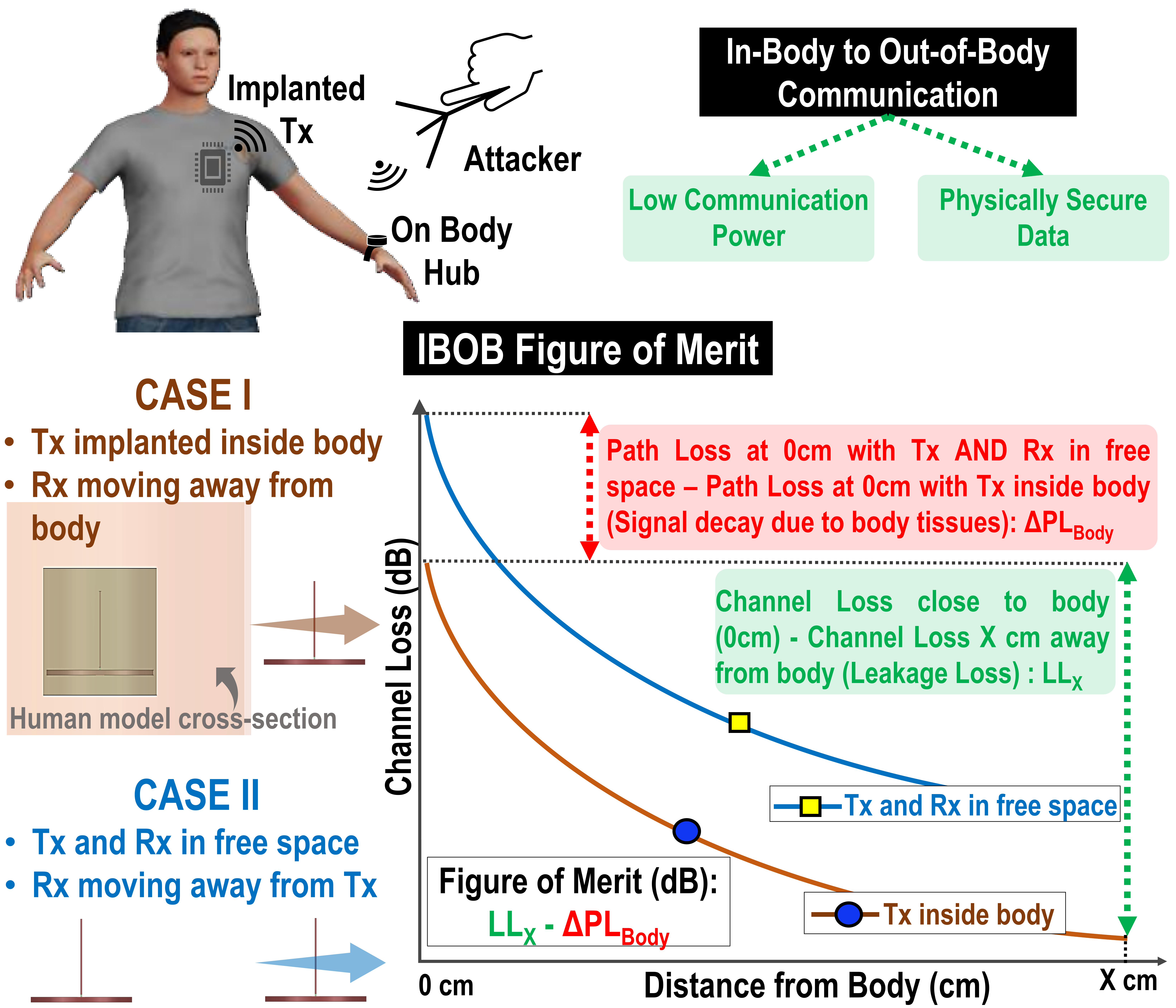}

\caption{In-Body to Out-of-Body communication is studied for various frequency bands and compared with respect to key parameters like communication power and physical layer security. A Figure of Merit is presented which encapsulates these key parameters to quantitatively compare different frequency bands.}
\vspace{-1em}
\label{fig:FOM}
\end{figure}

Smart devices in and around the body are rapidly becoming an integral part of our lifestyle, with applications such as wearable and implantable remote health monitoring devices redefining the healthcare sector. Unfortunately these developments also imply an abundantly present communication of sensitive data around ourselves \cite{NSR, nath2021inter}. Implantable Medical Devices like pacemakers and insulin pumps have been shown to be vulnerable to attacks resulting in fatal consequences (Fig. \ref{fig:FOM}) \cite{FDA, FDA2}. \newline
Thus, an in-depth study, of the in-body to out-of-body (IBOB) channel characteristics is essential in developing a communication architecture that is both efficient and secure. Previous works include development of channel models \cite{maity2018bio, datta2020advanced, modak2020bio, IMS2021, khan_wireless_2020} and efficient transceivers \cite{10.1145/3406238, maity2021sub, modak2022eqs, chatterjee20211, chatterjee202265nm, kiourti2012review} for IBOB communication. In this study, we present for the first time a quantitative study of physical layer security as a function of operating frequency using a Figure of Merit (FoM) (Fig. \ref{fig:FOM}) to assess the channel quality. A thorough analysis is performed through EM simulations and experiments of the effect of human body tissues on transmitted signals, and signal leakage away from the body for IBOB communication using frequently used ISM bands - 400 MHz, 900 MHz, 2.4 GHz and Human Body Communication (HBC) \cite{sen2020body,zimmerman1996personal, Wegmueller, Safety_Study} at 21 MHz. 



\renewcommand{\thefigure}{3}
\begin{figure*}[b!]
\centering

\includegraphics[width=0.95\textwidth]{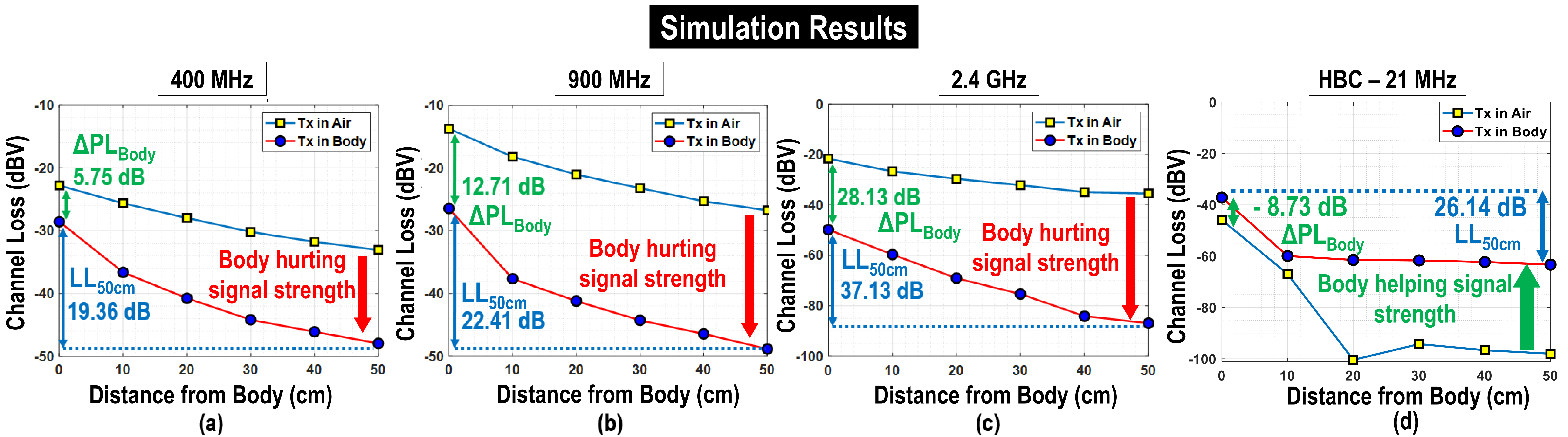}

\caption{Channel loss results vs distance from body as defined in Fig. 1 for FEM simulations at (a) 400 MHz, (b) 900 MHz, (c) 2.4 GHz, (d) HBC - 21 MHz.}

\label{fig:Sim_Results_RF}
\end{figure*}


\vspace{-0.5em}
\section{Theoretical Analysis and Simulations}
\subsection{Key Parameters for IBOB Communication}
Size-constrained implantable devices are required to have a low transmit power to ensure long device  usage. Further, critical information communicated from an implantable to a on-body hub needs to be protected to prevent eavesdropping by a skilled attacker. Thus, lowering the transmit power for longevity and minimizing signal leakage out of the body to ensure physical layer security, are the two key factors that help us define an FoM for secure and efficient IBOB communication. Further, the parameters defined must also be independent of the antenna or coupler parameters being used for the simulations and experiments to ensure that the FoM is strictly dependent on the physical properties of the signal and its interaction with the body tissues.
\vspace{-1em}
\subsection{Figure of Merit for IBOB Communication}
In the subsequent discussions, $PL_X$ refers to the path loss of the channel with the Rx at a distance $X$ away from body. 

\subsubsection{Physical Layer Security} 
To ensure physical layer security of the IBOB Communication channel, a high signal strength decay is required as we move away from the body. We define the parameter Leakage Loss at a distance $X$ away from the body $(LL_X)$ as a measure of signal decay as shown by (\ref{eqn:LL}).

\begin{equation}
        LL_X (dB) = PL_0 - PL_X
        \label{eqn:LL}
\end{equation}

This parameter $(LL_X)$ effectively captures the amount of signal decay as we move the Rx away from the body channel as illustrated by Fig. \ref{fig:FOM}. Higher the value of $LL_X$, more signal gets decayed away from body providing a more physically secure channel. 

\renewcommand{\thefigure}{2}
\begin{figure}[H]
\centering

\includegraphics[width=0.9\columnwidth]{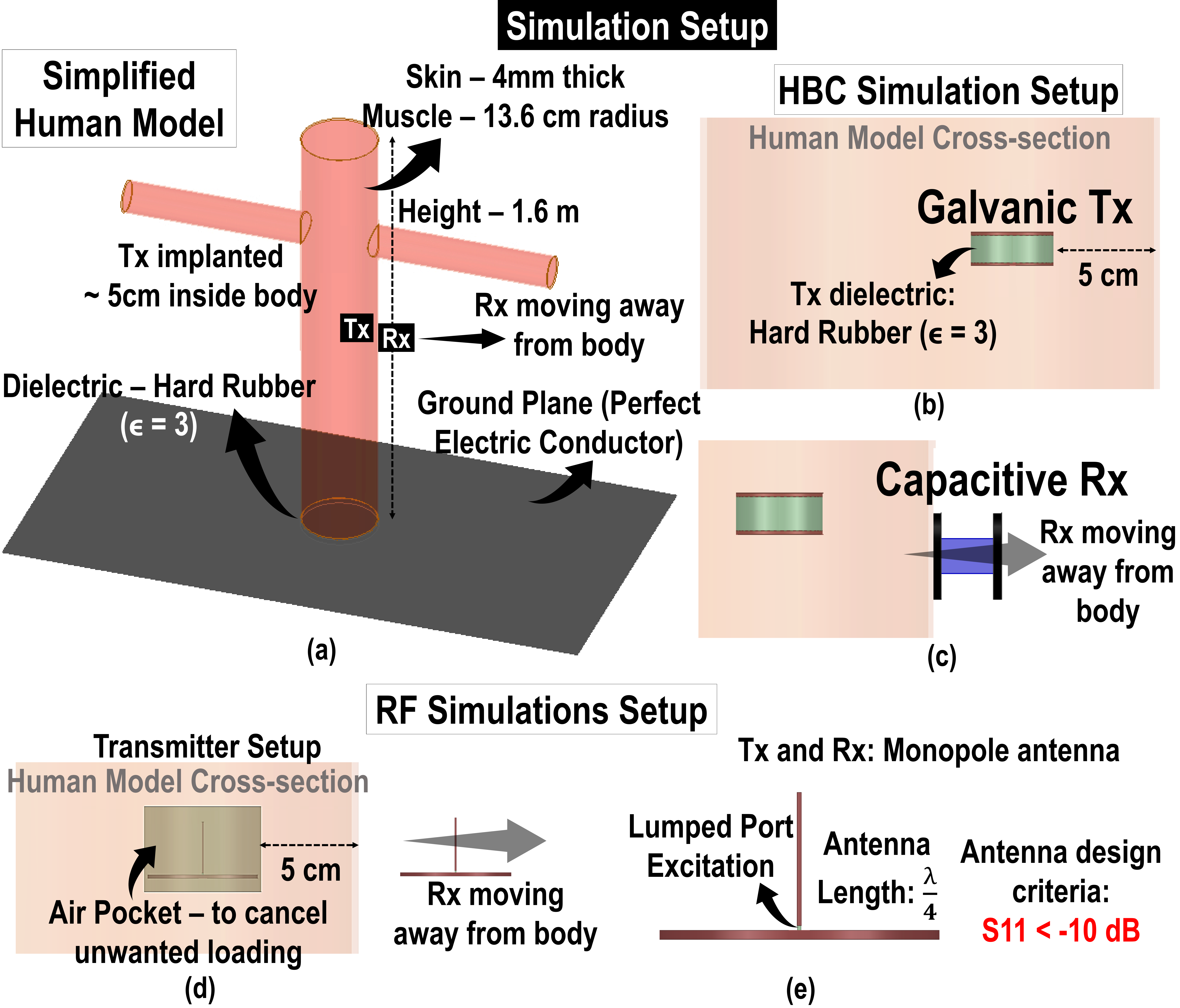}

\caption{(a) The simulations are performed using a simplified human model. The HBC simulations are done using (b) Galvanic Tx and (c) Capacitive Rx with the receiver moving away from the body. (d) RF simulations are performed with monopole antenna as Tx and Rx with the Tx embedded inside the body within an air pocket to cancel out unwanted loading effects. (e) Structure of Tx and Rx used for simulations.}

\label{fig:Sim_Setup}
\end{figure}


\subsubsection{Communication Power} Signal decay due to the body tissues is an essential factor in determining the transmit power of the system. Higher absorption of signal by body tissues implies we need a higher transmit power to ensure that the Rx sensitivity is enough for successful communication. As a measure of signal decay due to body tissues, we perform experiments and simulations in two different circumstances as shown in Fig. \ref{fig:FOM} (Case I and II). In case I, the Tx is present inside the body and the Rx is placed close to the body (0 cm away from body). The path loss in this case is termed as $PL_{0-Body}$. In case II, the Tx and Rx are placed at the same distance away from each other as in case I but without the presence of body or in other words, in free space. The path loss here is termed as $PL_{0-Air}$. The FoM parameter $\Delta PL_{Body}$ is defined as shown by (\ref{eqn:Del_PL}).

\begin{equation}
        \Delta PL_{Body} (dB) = PL_{0-Air} - PL_{0-Body}
        \label{eqn:Del_PL}
\end{equation}

The parameter $\Delta PL_{Body}$ captures the difference in channel loss that occurs due to the presence of body tissues affecting the transmitted signal. Hence, a low $\Delta PL_{Body}$ is desired for effective data transmission. A negative $\Delta PL_{Body}$ value indicates that the channel loss reduces due to the presence of body and the body in such a scenario helps the received signal quality instead of adding to channel loss. 

The FoM is defined by (\ref{eqn:FOM}).
\begin{equation}
        FoM (dB) = LL_X - \Delta PL_{Body}
        \label{eqn:FOM}
\end{equation}
Higher the value of FoM, the better a given frequency band is for IBOB communication. In this study, equal priority is given to both the parameters ($(LL_X)$, $\Delta PL_{Body}$). However, we may also use a weighted FoM to assign higher priority to one of the two parameters as per the requirements of the communication system.

\section{Simulations}

We perform Finite Element Method (FEM) based electromagnetic simulations on Ansys High Frequency Structure Simulator (HFSS) to validate the FoM defined. 
\subsection{Simulation setup}

A simplified crossed-cylindrical model of the human body made up of skin and muscle tissues is used for the simulations as shown in Fig. \ref{fig:Sim_Setup} (a). This simplified model is used to reduce computational complexity as well as simulation time. This simplified structure has been validated by comparing the EM field distribution around the model with that generated by a complex human model - VHP Female v2.2 by Neva Electromagnetics \cite{neva_model} which provided identical results. The simulations have been performed at 400 MHz, 900 MHz and 2.4 GHz which are part of the frequently used ISM bands as well as at 21 MHz which has been the standard defined by IEEE 802.15.6 for HBC. 

The Tx for HBC simulations (Fig. \ref{fig:Sim_Setup} (b)) is a galvanic mode voltage coupler \cite{IMS2021} which is embedded inside the human model. The Rx used is a capacitive mode voltage coupler which provides a low loss HBC channel at 21 MHz. The Rx is moved away from the body to measure the channel loss at various points as shown by Fig. \ref{fig:Sim_Setup} (c). 

A monopole antenna with lumped port excitation is used for RF simulations as the Tx and Rx to reduce design complexity as illustrated by Fig. \ref{fig:Sim_Setup} (d). The antenna was designed to have $S11 <-10dB$ to ensure efficient transmission of signals. However, the FoM calculations are independent of the antenna parameters and depend strongly on the signal and its interaction with the body tissues.\newline
Rx is moved away from Tx for two cases described in section II.B.2: 1) Tx inside the body, 2) Tx in air or free space. 

\subsection{Simulation Results}

The simulation results are illustrated in Fig. \ref{fig:Sim_Results_RF}. The values of the critical parameters $((LL_X)$, $\Delta PL_{Body})$ are highlighted in the figures. The simulation results show how the signal decays away from the body for the different communication methodologies. We observe that for HBC (Fig. \ref{fig:Sim_Results_RF} (d)), the effect of body on the signal changes as the body starts helping the channel loss instead of hurting it by acting like a wire to provide a low loss channel as compared to when no body is present. Thus, as explained in Section II.B.2, the value of $\Delta PL_{Body}$ becomes negative. Higher conductivity of body tissues for larger frequencies (2.4 GHz) results in a higher attenuation inside the body thus degrading the FoM. A comparison of the FoM described by (\ref{eqn:FOM}) is provided by the table in Fig: \ref{table:FOM_Sims}. The table shows that FoM for HBC is at least an order of magnitude higher than that observed for RF communication.

\renewcommand{\thefigure}{4}
\begin{figure}[t!]
\centering
\includegraphics[width=0.7\columnwidth]{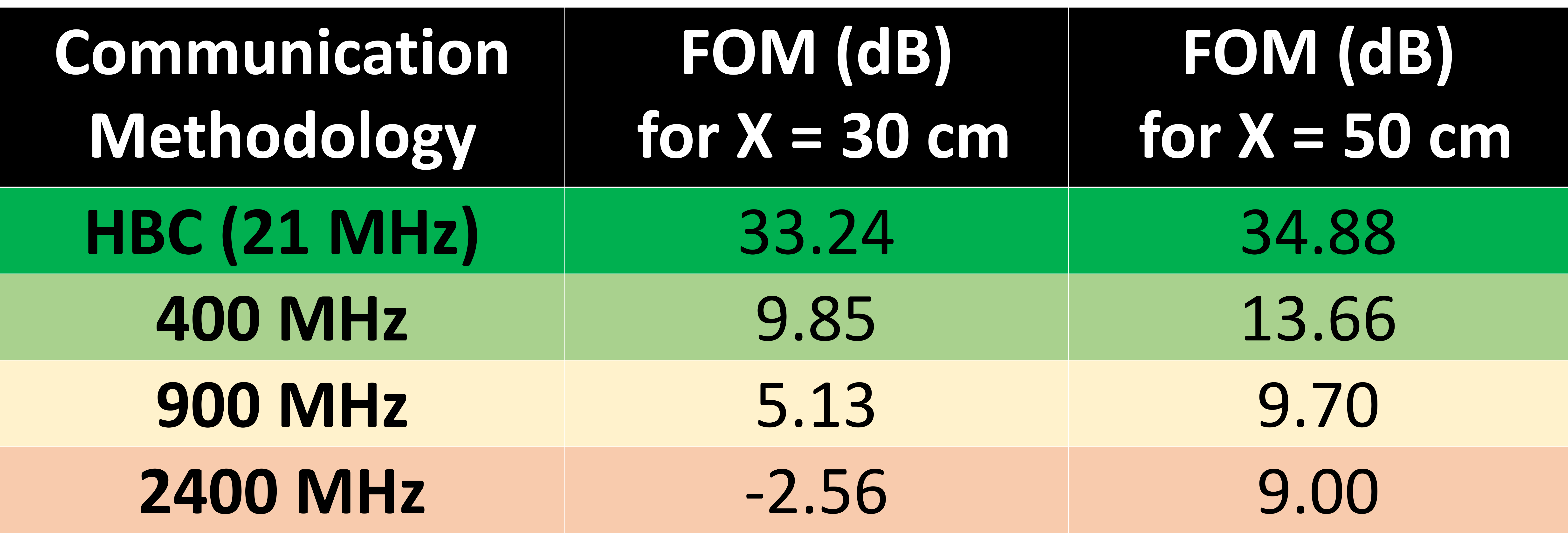}
\caption{Figure of Merit calculated for the FEM simulations performed.}

\label{table:FOM_Sims}
\end{figure}

\renewcommand{\thefigure}{5}
\begin{figure}[b!]
\centering

\includegraphics[width=0.47\textwidth]{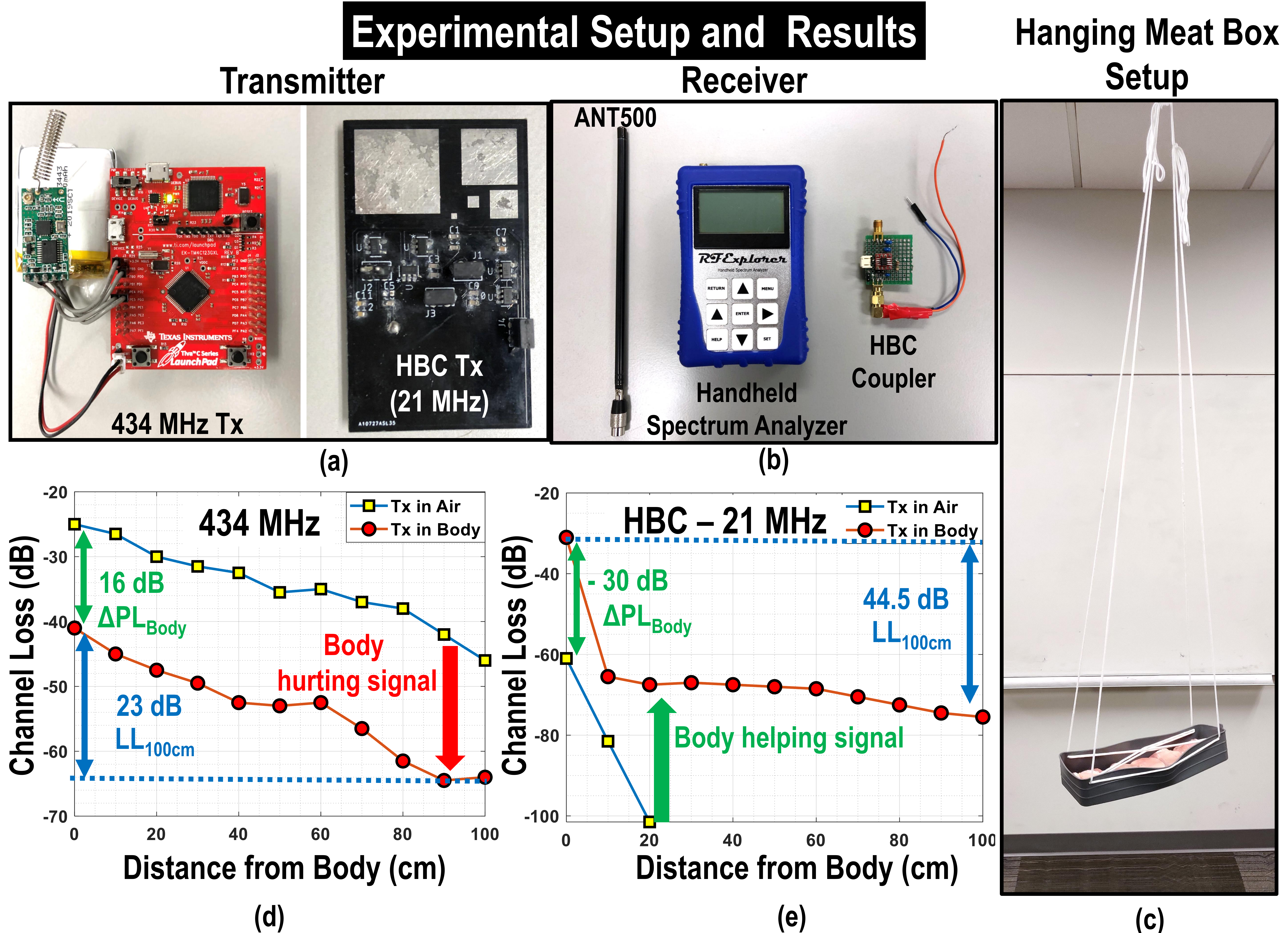}

\caption{(a) Tx and (b) Rx used for 434 MHz RF communication and 21 MHz HBC. (c) Hanging meat box setup used to perform experiments. Channel loss results obtained for (d) 434 MHz, (e) 21 MHz.}

\label{fig:Expt_Results}
\end{figure}

\section{Experiments}

To further validate the observed trends shown by the simulations, we perform experiments for HBC at 21 MHz and compare it with RF communication at 434 MHz as the 400 MHz band provided us with the best FoM for RF IBOB communication.  

\subsection{Experimental setup}

The experiments are performed in a standard lab environment with a hanging meat box setup (Fig. \ref{fig:Expt_Results} (c)) to model the human body. A hanging setup is essential in accurately observing the HBC results. \cite{IMS2021} The galvanic transmitter for HBC is a 3/5 stage ring oscillator on FR4 PCB transmitting at 21 MHz as shown in Fig. \ref{fig:Expt_Results} (a). The RF transmitter (Fig. \ref{fig:Expt_Results} (a)) is designed using a 434 MHz transceiver module by Cytron Technologies. The transmitter is controlled using a TIVA C Launchpad microcontroller using UART serial communication protocol. The transmitters are embedded inside layers of pork meat to emulate an implanted device. Pork meat is used due to the close resemblance in their dielectric properties with human tissues. The receiver (Fig. \ref{fig:Expt_Results} (b)) is an RF Explorer handheld Spectrum Analyzer. For the HBC measurements, we use a coupler with a high impedance termination obtained using a broadband buffer - BUF602ID from Texas Instruments (Fig. \ref{fig:Expt_Results} (b)). The RF signals are obtained using an ANT500 antenna by Great Scott Gadgets. The path loss is measured as we move away from the meat box setup with the Tx inside the meat layers (case I) and in air (case II) to observe the decay in signal strength.
\vspace{-0.5em}
\subsection{Experimental Results} 

The experimental results are illustrated in Fig. \ref{fig:Expt_Results} (d) and (e). We observe that the signal decay in HBC (Fig. \ref{fig:Expt_Results} (e)) away from the body is much higher than that observed for RF transmission at 434 MHz (Fig. \ref{fig:Expt_Results} (d)). Further, the FoM for HBC and 434 MHz RF communication is compared in the table in Fig. \ref{table:FOM_Expts}. The difference in FoM values observed in simulation and experiments rises from the variation in experimental and simulation conditions. The experiments conducted in a lab environment results in higher attenuation of signals resulting in a faster decay. Further, the thin layer of meat used in the experiments allows for high dipole coupling between Tx and Rx resulting in a low path loss for HBC measurements with Rx close to the Tx thus further increasing the measure of signal decay. However, the trends shown in the experimental results as well as the FoM values match with the simulation results and the FoM trends where HBC is observed to provide a much better performance than RF communication techniques for IBOB communication. 
\renewcommand{\thefigure}{6}
\begin{figure}[t!]
\centering
\includegraphics[width=0.65\columnwidth]{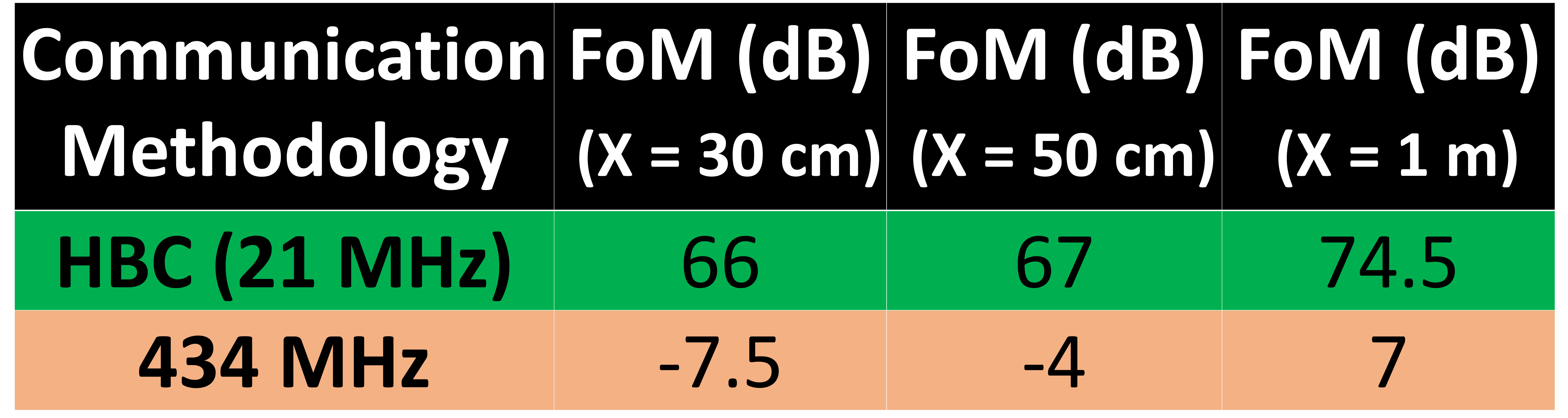}
\caption{FoM for the experiments performed at 21 MHz (HBC), 434 MHz.}

\label{table:FOM_Expts}
\end{figure}

\section{Conclusion}

A quantitative study for physical layer security in IBOB communication as a function of frequency is presented for the first time in literature. A FoM is developed to compare the performance of the IBOB channel for the different frequencies which shows that HBC operating at 21 MHz provides order(s) of magnitude better performance for IBOB communication compared to typical RF based communication methodologies. 


\bibliographystyle{IEEEtran}

\bibliography{references.bib}

\end{document}